\documentclass[prl,aps,twocolumn,showpacs]{revtex4}
\usepackage{epsf}
\usepackage{amsmath}

\begin{document}
\title{Efficient generation of distant atom entanglement}

\author{Grzegorz Chimczak}

\affiliation{Nonlinear Optics Division, Physics Institute, Adam
 Mickiewicz University, 61-614 Pozna\'n, Poland}

\date{\today}
\email{chimczak@kielich.amu.edu.pl}

\begin{abstract}
We show how the entanglement of two atoms, trapped in distant
separate cavities, can be generated with arbitrarily high probability of
success. The scheme proposed employs sudden excitation of the atoms proving that
the weakly driven condition is not necessary to obtain the success rate
close to unity. The modified scheme works properly even if each cavity
contains many atoms interacting with the cavity modes. We also show
that our method is robust against the spontaneous atomic decay.
\end{abstract}

\maketitle

\thispagestyle{empty}

Many quantum information tasks require an entanglement,
especially an entanglement shared by distant atoms can play a very
important role in quantum information processing. This is due
to the fact that atomic states are ideal for quantum
information storage. Therefore, a variety of schemes for entanglement of
distant atoms have been proposed recently~\cite{duan_nature,cabrillo99,bose,
browne_entanglement,feng_entanglement,duan:_effic,simon_entanglement,
clark_entanglement,zou_4distant}.
The schemes employ also photonic states providing fast
quantum information transfer over long distances.
Most of the schemes describe two cavities, each containing
one trapped atom. The photons leaking out from the cavities are
mixed at a beam splitter and detected~\cite{bose,browne_entanglement,
feng_entanglement,duan:_effic,simon_entanglement}. 
In those schemes, however, only two of the four Bell states can be used,
and therefore, the success rate is less than $50\%$~\cite{lutkenhaus,vitali}.
Moreover, the success rate is lowered by the spontaneous atomic
emissions. In most of the schemes the population of the excited
state is considerable~\cite{feng_entanglement,duan:_effic,simon_entanglement}
during the entangling operations
and can therefore drastically lower the success rate as it
has been proved in~\cite{chimczak02:_effect}.
Only the proposal of Browne \emph{et~al}.~\cite{browne_entanglement} avoids
all of the above problems.
The whole operation is performed there in such a way that the population
of the excited state is negligible thanks to the use of large
detunings. Furthermore, the scheme uses the requirement of weak driving since
a sudden excitation of the atoms limits the entanglement efficiency to $50\%$ as is
suggested in~\cite{browne_entanglement}. However, the condition makes
it impossible to
perform many operation requiring strong driving, and therefore,
it can be difficult to use the entangled atoms in quantum computations.
It is possible to change this condition by controlling laser intensity
but then the entanglement operation time will be long.

\begin{figure}[htbp]
  \begin{center}
\epsfxsize=7.0cm\epsfbox{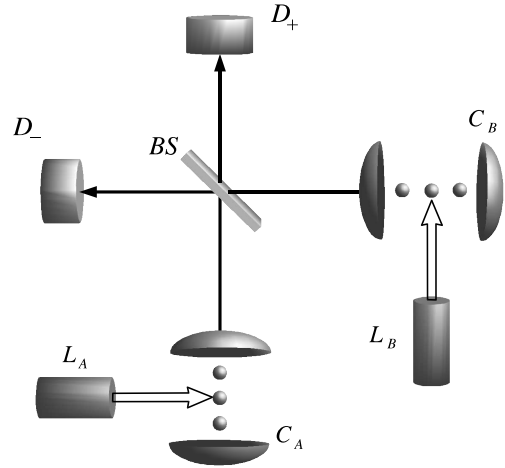} \caption{Schematic
representation of the setup to generate entangled state of two distant atoms. }
\label{fig:rys_uklad}
  \end{center}
\end{figure}

\begin{figure}[htbp]
  \begin{center}
\epsfxsize=7.0cm\epsfbox{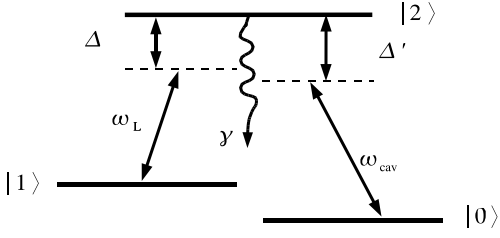}  \caption{Level scheme of one
of the $\Lambda$ atoms interacting with the classical laser field
and the quantized cavity mode.}
 \label{fig:rys_poziomy}
  \end{center}
\end{figure}

In this paper a scheme is presented that employs sudden excitation
to entangle two atoms with high success probability.
The main idea of the scheme is to use a protocol which first prepares each
cavity in one photon state, next creates maximally entangled state of both
cavity fields detecting one photon decay from the cavities and
finally maps the entangled state onto two distant atoms.
The strong driving condition makes it possible to use the entangled
distant atoms in various quantum information tasks.
The setup consists of two cavities, a $50$-$50$ beam splitter,
two lasers ($L_{A}$ and $L_{B}$) and two detectors ($D_{+}$ and $D_{-}$)
as depicted in figure~\ref{fig:rys_uklad}.
Quantum computations require also many
qubits~\cite{jane02,chimczak:_entanglement_teleportation} thus each cavity
can contain up to $N$ atoms. Each atom is modeled by a three-level
$\Lambda$ system with one excited state $|2\rangle$ and two
ground states $|0\rangle$ and $|1\rangle$. The energy level
structure of the atom is shown in figure~\ref{fig:rys_poziomy}.
The spontaneous decay
rate of the excited state is denoted by $\gamma$. There are
two transitions in the $\Lambda$-type atom. First of them,
the $|1\rangle \leftrightarrow |2\rangle$ transition, is driven by classical
laser field with the coupling strength $\Omega$. The frequency
of the laser field is $\omega_{\text{L}}$.
The second, the $|0\rangle \leftrightarrow |2\rangle$ transition, is coupled
to the cavity mode with a frequency $\omega_{\text{cav}}$ and coupling
strength $g$. Both the classical laser field and the quantized cavity mode
are detuned from the corresponding transitions.
The two detunings are given by
$\Delta=(E_{2}-E_{1})/\hbar-\omega_{\text{L}}$ and
$\Delta^{\prime}=(E_{2}-E_{0})/\hbar-\omega_{\text{cav}}$.
The evolution of the quantum system is conditional.
During the time intervals when no photon decay is detected, the
evolution is governed by the effective non-Hermitian Hamiltonian
($\hbar=1$ here and in the following)
\begin{eqnarray}
  \label{eq:Hamiltonian0}
  H&=&\sum_{k} (\Delta - i \gamma) \sigma_{22}^{(k)}
  - \sum_{k} \Delta_{r} \sigma_{00}^{(k)}-i \kappa
  a^{\dagger} a \nonumber \\
  &&+\sum_{k} (\Omega \sigma_{21}^{(k)}+g a \sigma_{20}^{(k)}+ {\rm{H.c.}}) \, ,
\end{eqnarray}
where $a$ denotes the annihilation operator for the cavity mode,
$\kappa$ is the cavity decay rate,
$\Delta_{r}=\Delta^{\prime}-\Delta$  and $k$ indicates the atom.
In expression~(\ref{eq:Hamiltonian0}) the flip operators
$\sigma_{ij}^{(k)}\equiv (|i\rangle \langle j|)_{k}$, where
$i,j=0,1,2$, are also used.
The evolution generated by~(\ref{eq:Hamiltonian0}) is interrupted by
collapses corresponding to the action of the operator
\begin{equation}
  \label{eq:operatorColl}
  C=\sqrt{\kappa} (a_{A}+\epsilon a_{B}) \, ,
\end{equation}
where $a_{A}$ and $a_{B}$ denote the annihilation operators for $A$ and
$B$ cavity modes, respectively, and $\epsilon$ is equal to unity
for the photon detection in $D_{+}$ and minus unity
for the photon detection in $D_{-}$. The Hamiltonian~(\ref{eq:Hamiltonian0})
takes a simpler form in the large detunings limit
($\Delta \gg \Omega$ and $\Delta^{\prime} \gg g$), when the excited state can be
adiabatically eliminated~\cite{pell,alexanian95,carmichaelKsiazkaMethods}.
In order to avoid the lowering of the success probability by
spontaneous atomic decay, it is necessary to assume that
$\gamma \ll \Delta, \Delta^{\prime}$ and
$\gamma g^{2}/{\Delta^{\prime}}^{2} ,
\gamma \Omega^{2}/{\Delta}^{2} \ll \kappa$~\cite{chimczak02:_effect}.
Then the Hamiltonian~(\ref{eq:Hamiltonian0}) passes into
\begin{eqnarray}
\label{eq:Hamil1}
H&=&-\sum_{k} (\Delta_{r} \sigma_{00}^{(k)}
+z_{1} \sigma_{11}^{(k)}+z_{2} a^{\dagger} a \sigma_{00}^{(k)})
\nonumber   \\
& &-\sum_{k}(z_{3} a \sigma_{10}^{(k)} +{\rm{H.c.}})
-i \kappa a^{\dagger} a \, ,
\end{eqnarray}
where $z_{1}= \Omega^{2} / \Delta$, $z_{2}=g^{2}/\Delta^{\prime}$ and
$z_{3}=\frac{1}{2} \Omega g ({\Delta^{\prime}}^{-1}+\Delta^{-1})$.
As mentioned above, it would be very useful to entangle two distant
atoms within the strong driving limit. Therefore, we assume
$z_{3} \gg \kappa$.
In order to further simplify the problem we now assume that there
is only one atom inside each cavity.
This allows us to assume that $\Omega=g$ and $\Delta_{r}=0$.
We will later extend the model over the case of many atoms.
Under the conditions, the Hamiltonian~(\ref{eq:Hamil1}) takes the
form
\begin{eqnarray}
\label{eq:Hamil1b}
H&=&-z \sigma_{11}-z a^{\dagger} a \sigma_{00}
-(z a \sigma_{10} +{\rm{H.c.}})
-i \kappa a^{\dagger} a \, . \nonumber
\end{eqnarray}
The protocol needs two local operations. The first of them is to map the 
atomic state onto the cavity mode and the second of them is to map the
photonic state in the atom.
The local operations necessary to achieve the entanglement can be
obtained via $e^{-i H t}$. Let us denote by
$|jn\rangle \equiv |j\rangle \otimes |n\rangle$ a state of
the system consisting of one atom
in the state $|j\rangle$ trapped inside a cavity with $n$ photons.
One can perform the two local operations
by illuminating the atom for times $t_{1}$ and $t_{2}$
\begin{eqnarray}
\label{eq:U1}
|10\rangle &\rightarrow &
i e^{i z t_{1}} e^{-\frac{\kappa t_{1}}{2}} |01\rangle \, , \nonumber \\
\label{eq:D1}
|01\rangle &\rightarrow &
i e^{i z t_{2}} e^{-\frac{\kappa t_{2}}{2}} |10\rangle \nonumber \, ,
\end{eqnarray}
where $t_{1}=\frac{2}{\Omega_{\kappa}}
\big[ \pi-\arctan(\frac{\Omega_{\kappa}}{\kappa})\big]$,
$t_{2}=\frac{2}{\Omega_{\kappa}} \arctan(\frac{\Omega_{\kappa}}{\kappa})$
and $\Omega_{\kappa}=\sqrt{4 z^{2}-\kappa^{2}}$.
When the laser is turned off, the system's evolution is given by
\begin{eqnarray}
\label{eq:las0a}
e^{-i H t} |10\rangle&=& |10\rangle \, , \nonumber \\
\label{eq:las0b}
e^{-i H t} |01\rangle&=& e^{i z t} e^{-\kappa t}|01\rangle  \, .
\end{eqnarray}
The state $|00\rangle$ always remains unchanged even if the laser is
turned on.

At the beginning of the protocol, both cavity fields are empty and
both atoms are prepared in the ground state $|1\rangle$. Thus,
the initial state is given by $|10\rangle_{A}$ $\otimes$ $|10\rangle_{B}$.
The protocol consists of four stages.

(i) In the first stage of the protocol the two atoms are illuminated by
the lasers for the time $t_{1}$. On condition that no photon detection
occurs during the time, the state becomes 
$|01\rangle_{A}$ $\otimes$ $|01\rangle_{B}$.
Then, one should begin the detection stage which is the
second stage of the protocol.
However, if one collapse has been detected during the illuminating time $t_{1}$,
the jump operator $C$~(\ref{eq:operatorColl}) acts on the state.
In this case, we obtain the entangled state of both cavity fields
$(|00\rangle_{A} |01\rangle_{B}+\epsilon |01\rangle_{A} |00\rangle_{B})$
and the detection stage is superfluous. It means that the third
stage of the protocol should be started.
The detection of two photons at the stage means that the entanglement
procedure has failed.

(ii) In the second stage, one waits until the click is recorded in either
of the detectors. During the detection time the lasers are turned off
and the evolution is given by~(\ref{eq:las0b}).
Detection of one photon corresponds to the action of the jump operator
$C$~(\ref{eq:operatorColl}) and thus the state becomes
$(|00\rangle_{A} |01\rangle_{B}+\epsilon |01\rangle_{A} |00\rangle_{B})$.
After the detection event, the third stage of the protocol should be
started immediately.

(iii) The third stage is responsible for mapping and storage of the entangled
state of both cavity fields in the state of the two atoms. This can be
realized by turning the lasers on for the period of time $t_{2}$.
After this operation the state is given by
$(|00\rangle_{A} |10\rangle_{B} +\epsilon |10\rangle_{A} |00\rangle_{B})$.
If any collapse is detected during this stage the entangling process is
unsuccessful.

(iv) If $D_{+}$ clicks, the whole protocol is over because $\epsilon=1$.
However, if $D_{-}$ clicks one has to remove the phase shift factor
using the Zeeman evolution. This is the objective of the fourth stage.

\begin{figure}[htbp]
  \begin{center}
     \epsfxsize=7.0cm\epsfbox{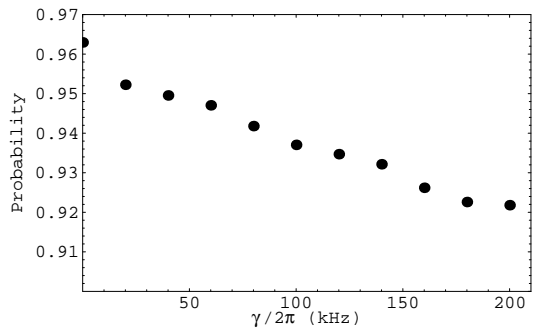}
    \caption{The average probability of success calculated numerically
    as a function the spontaneous decay rate. The averages are
    taken over 20 000 trajectories. The parameter regime is
    ($\Delta$; $\Delta^{\prime}$; $\Omega$; $g$; $\kappa$)$/2 \pi=$
      ($300$; $300$; $25$; $25$; $0.05$) MHz.}
    \label{fig:rysP_gamma.eps}
  \end{center}
\end{figure}

\begin{figure}[htbp]
  \begin{center}
     \epsfxsize=7.0cm\epsfbox{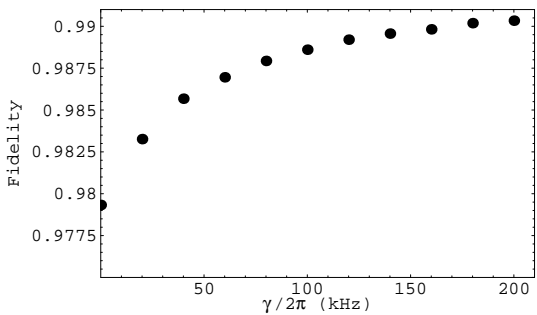}
    \caption{The entanglement fidelity computed numerically 
    as a function the spontaneous decay rate for
    ($\Delta$; $\Delta^{\prime}$; $\Omega$; $g$; $\kappa$)$/2 \pi=$
    ($300$; $300$; $25$; $25$; $0.05$) MHz.}
    \label{fig:rysF_gamma.eps}
  \end{center}
\end{figure}

Finally, the entangled state of the two distant atoms is obtained. The
probability that the protocol will be successful, under the strong driving 
condition, can be well approximated by
\begin{eqnarray}
\label{eq:Psuc}
P_{suc}&=&e^{-\alpha \pi} (2-e^{-\alpha \frac{\pi}{2}}) \, ,
\end{eqnarray}
where $\alpha=\kappa / z$.
\begin{figure}[htbp]
  \begin{center}
     \epsfxsize=7.0cm\epsfbox{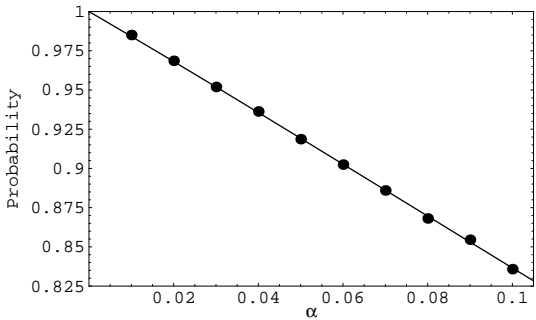}
    \caption{The average probability of success given by~(\ref{eq:Psuc})
    (solid curve) and computed numerically (points) for parameters
    ($\Delta$; $\Delta^{\prime}$; $\Omega$; $g$; $\gamma$)$/2 \pi=$
      ($300$; $300$; $25$; $25$; $0$) MHz. }
    \label{fig:rysP_alpha.eps}
  \end{center}
\end{figure}
Figure~\ref{fig:rysP_alpha.eps} shows that the
probability of success tends to unity with decreasing $\alpha$.
For reference, we also calculate numerically the probability of success
using the quantum trajectory theory~\cite{carmichaelksiazka,pleniotrajek}.
We use the non-Hermitian
Hamiltonian~(\ref{eq:Hamiltonian0}) in all numerical computations.
The parameters ($\Delta$; $\Delta^{\prime}$; $\Omega$; $g$; $\kappa$; $\gamma$)$/2 \pi=$
($300$; $300$; $25$; $25$; $0.05$; $0.1$) MHz are chosen is such a way that all
the aforementioned assumptions ($\Delta_{r}=0$; $\Delta \gg \gamma$;
$\Omega=g$; $\Delta \gg \Omega$;
$z \gg \kappa \gg \gamma \Omega^{2}/{\Delta}^{2}$) are
satisfied. We find that the average fidelity of the entanglement is
about $0.99$ and the average probability of success is about $0.94$.
The averages are taken over twenty thousand trajectories.
Moreover, we set $\gamma$ to zero to verify the analytical expression
describing the success rate given by~(\ref{eq:Psuc}).
In figure~\ref{fig:rysP_alpha.eps}, the points show the average
probability of success over twenty thousand trajectories
calculated for different values of $\kappa$. As one can see,
the analytical results are in a remarkable agreement with the
numerical solution.
We have also investigated the influence of the spontaneous decay rate
of the excited state on the entanglement. We plot in
figure~\ref{fig:rysP_gamma.eps} the average probability of success
as a function of the spontaneous decay rate.
As evident from the figure, the probability of success decreases
with increasing $\gamma$. Figure~\ref{fig:rysF_gamma.eps} shows
the influence of the decay rate on the entanglement fidelity.
Surprisingly, an increase in the spontaneous decay rate leads to
an improvement in the fidelity. This is due to the fact that
the fidelity depends on the population of the excited state.
If saturation parameters ($g^{2}/{\Delta^{\prime}}^{2}$,
$\Omega^{2}/{\Delta}^{2}$)
are not small enough, the population can be lowered
by a sufficiently high spontaneous decay rate.

Let us now consider the case of many atoms present in each cavity.
All atoms, except the two distant ones that are to be entangled,
can be in arbitrary states. Unfortunately, all atoms
that are in state $|0\rangle$ interact with the cavity mode. This
makes it impossible to perform any operation on a separate atom without
changing the other atoms' states.
One can avoid the problem using the $z_{3} \gg z_{2}$ condition,
which is equivalent to the assumption $\Omega \gg g$. It is also
necessary to assume that $\Delta_{r}=z_{1}$.
In the above limits, the evolution generated by~(\ref{eq:Hamil1}) can
be used to obtain the necessary local operations. Let us denote
the state of all atoms that are not intended to be entangled with
a distant atom by $|\Phi\rangle$. We assume that
a number of the atoms in state $|0\rangle$ which do not
participate in the entanglement process is $N_{0}$.
We can approximately perform a quantum state mapping by illuminating
the atom for the time $t_{3}=\pi /(2 z_{3})$:
\begin{eqnarray}
\label{eq:UOmwg}
|\Phi\rangle |10\rangle &\rightarrow &
i e^{i \Delta_{r} (N_{0}+1) t_{3}} e^{-\frac{\kappa t_{3}}{2}}
e^{i \frac{N_{0}+1}{2} z_{2} t_{3}}
|\Phi\rangle |01\rangle \, , \nonumber \\
\label{eq:DOmwg}
|\Phi\rangle |01\rangle &\rightarrow &
i e^{i \Delta_{r} (N_{0}+1) t_{3}} e^{-\frac{\kappa t_{3}}{2}}
e^{i \frac{N_{0}+1}{2} z_{2} t_{3}}
|\Phi\rangle |10\rangle \nonumber \, .
\end{eqnarray}
When the laser is turned off, the evolution of the states is given by
\begin{eqnarray}
\label{eq:BezLasOmwg0}
e^{-i H t} |\Phi\rangle |10\rangle &=& e^{i \Delta_{r} N_{0} t}
|\Phi\rangle |10\rangle \, , \nonumber \\
\label{eq:BezLasOmwg1}
e^{-i H t} |\Phi\rangle |01\rangle &=&
e^{i (\Delta_{r} +z_{2}) (N_{0}+1) t} e^{-\kappa t}
|\Phi\rangle |01\rangle \nonumber \, .
\end{eqnarray}
If the atom is in the ground state $|0\rangle$ and the cavity field is empty,
the population of the state always remains unchanged (even if laser is turned
on)
\begin{eqnarray}
\label{eq:Omwg00}
e^{-i H t} |\Phi\rangle |00\rangle &=& e^{i \Delta_{r} (N_{0}+1) t}
|\Phi\rangle |00\rangle \, . \nonumber
\end{eqnarray}

Now let us present a modified scheme. The initial state is given by
$|\Phi\rangle_{A} |10\rangle_{A}$ $\otimes$ $|\Phi\rangle_{B} |10\rangle_{B}$.
The new protocol consists of three steps.

(i) Illumination of chosen distant atoms (one from cavity $A$ and
one from cavity $B$) for the time $t_{3}$. On condition that
no photon decay has been recorded, the state becomes
$|\Phi\rangle_{A} |01\rangle_{A}$ $\otimes$ $|\Phi\rangle_{B} |01\rangle_{B}$.
If one photon has been detected in time $t_{j}$, the state is
$|\Phi\rangle_{A} |\Phi\rangle_{B} (|00\rangle_{A} |01\rangle_{B}$
$+\epsilon \theta(t_{3}-t_{j}) |01\rangle_{A} |00\rangle_{B})$
where $\theta(t)=\exp[\frac{i}{2} (N_{0A}-N_{0B}) z_{2} t]$.
In this case the third stage should be started.

(ii) Waiting for one photon decay. After detection the state
becomes $|\Phi\rangle_{A} |\Phi\rangle_{B} (|00\rangle_{A} |01\rangle_{B}
+\epsilon |01\rangle_{A} |00\rangle_{B})$.

(iii) Mapping and storage of the entangled state of both cavity
fields in the two distant atoms by illuminating them for the time $t_{3}$.
The illuminating operations do not start simultaneously.
The laser $L_{B}$ is turned on after a delay $t_{\phi}$.
The state is then given by
$|\Phi\rangle_{A} |\Phi\rangle_{B} (|00\rangle_{A} |10\rangle_{B}
+\phi |10\rangle_{A} |00\rangle_{B})$,
where $\phi=\epsilon \theta(2 t_{3}-t_{j}) \exp[-i \Delta_{r} t_{\phi}]$
if photon decay has registered in the first stage and
$\phi=\epsilon \theta(t_{3}) \exp[-i \Delta_{r} t_{\phi}]$ for
photon detection in the second stage. We choose such a time $t_{\phi}$
that $\phi=1$.

The probability of success of the modified protocol is given
by expression~(\ref{eq:Psuc}) with $\alpha=\kappa / z_{3}$.
Finally, we have performed the numerical calculations for the parameters
($\Delta$; $\Delta^{\prime}$; $\Omega$; $g$; $\kappa$; $\gamma$)$/2 \pi=$
($1000$; $1000.9$; $30$; $0.7$; $0.001$; $0.1$) MHz which satisfy all the above
assumptions ($\Delta_{r}=z_{1}$;
$\Delta \gg \Omega \gg g$; $\Delta \gg \gamma$;
$z_{3} \gg \kappa \gg \gamma \Omega^{2}/{\Delta}^{2} $).
We assume that there are three atoms inside each cavity as in
figure~\ref{fig:rys_uklad}. We have generated one thousand
trajectories to compute the average of the probability of
success and the average of the fidelity.
For each trajectory we have generated random numbers
$N_{0A}$, $N_{0B} \in \{0,1,2\}$ and corresponding to them random
initial states $|\Phi\rangle_{A}$,
$|\Phi\rangle_{B} \in \{ |1\rangle_{1} |1\rangle_{2}$,
$c_{1} |1\rangle_{1} |0\rangle_{2} + c_{2} |0\rangle_{1} |1\rangle_{2}$,
$|0\rangle_{1} |0\rangle_{2} \}$, where $c_{1}$ and $c_{2}$ are arbitrary
complex amplitudes. We have obtained
the fidelity of $0.99$ and the success rate of $0.90$.

In conclusion, we have presented a scheme to create an entangled state of two
distant atoms. We have shown that the probability of success can be made
arbitrarily close to unity without the weakly driven condition.
The scheme works properly even if there are many atoms inside each
cavity. We have also investigated the influence of the spontaneous
decay rate of the excited state on the entanglement and we have found
that with the increasing rate the probability of success is slightly lower
and the entanglement fidelity is better.

%\bibliography{qinformation}

\begin{thebibliography}{19}
\expandafter\ifx\csname natexlab\endcsname\relax\def\natexlab#1{#1}\fi
\expandafter\ifx\csname bibnamefont\endcsname\relax
  \def\bibnamefont#1{#1}\fi
\expandafter\ifx\csname bibfnamefont\endcsname\relax
  \def\bibfnamefont#1{#1}\fi
\expandafter\ifx\csname citenamefont\endcsname\relax
  \def\citenamefont#1{#1}\fi
\expandafter\ifx\csname url\endcsname\relax
  \def\url#1{\texttt{#1}}\fi
\expandafter\ifx\csname urlprefix\endcsname\relax\def\urlprefix{URL }\fi
\providecommand{\bibinfo}[2]{#2}
\providecommand{\eprint}[2][]{\url{#2}}

\bibitem[{\citenamefont{Duan et~al.}(2001)\citenamefont{Duan, Lukin, Cirac, and
  Zoller}}]{duan_nature}
\bibinfo{author}{\bibfnamefont{L.~M.} \bibnamefont{Duan}},
  \bibinfo{author}{\bibfnamefont{M.~D.} \bibnamefont{Lukin}},
  \bibinfo{author}{\bibfnamefont{J.~I.} \bibnamefont{Cirac}}, \bibnamefont{and}
  \bibinfo{author}{\bibfnamefont{P.}~\bibnamefont{Zoller}},
  \bibinfo{journal}{Nature} \textbf{\bibinfo{volume}{414}},
  \bibinfo{pages}{413} (\bibinfo{year}{2001}).

\bibitem[{\citenamefont{Cabrillo et~al.}(1999)\citenamefont{Cabrillo, Cirac,
  Garc\'ia-Fern\'andez, and Zoller}}]{cabrillo99}
\bibinfo{author}{\bibfnamefont{C.}~\bibnamefont{Cabrillo}},
  \bibinfo{author}{\bibfnamefont{J.~I.} \bibnamefont{Cirac}},
  \bibinfo{author}{\bibfnamefont{P.}~\bibnamefont{Garc\'ia-Fern\'andez}},
  \bibnamefont{and} \bibinfo{author}{\bibfnamefont{P.}~\bibnamefont{Zoller}},
  \bibinfo{journal}{Phys. Rev. A} \textbf{\bibinfo{volume}{59}},
  \bibinfo{pages}{1025} (\bibinfo{year}{1999}).

\bibitem[{\citenamefont{Bose et~al.}(1999)\citenamefont{Bose, Knight, Plenio,
  and Vedral}}]{bose}
\bibinfo{author}{\bibfnamefont{S.}~\bibnamefont{Bose}},
  \bibinfo{author}{\bibfnamefont{P.~L.} \bibnamefont{Knight}},
  \bibinfo{author}{\bibfnamefont{M.~B.} \bibnamefont{Plenio}},
  \bibnamefont{and} \bibinfo{author}{\bibfnamefont{V.}~\bibnamefont{Vedral}},
  \bibinfo{journal}{Phys. Rev. Lett.} \textbf{\bibinfo{volume}{83}},
  \bibinfo{pages}{5158} (\bibinfo{year}{1999}).

\bibitem[{\citenamefont{Browne et~al.}(2003)\citenamefont{Browne, Plenio, and
  Huelga}}]{browne_entanglement}
\bibinfo{author}{\bibfnamefont{D.~E.} \bibnamefont{Browne}},
  \bibinfo{author}{\bibfnamefont{M.~B.} \bibnamefont{Plenio}},
  \bibnamefont{and} \bibinfo{author}{\bibfnamefont{S.~F.}
  \bibnamefont{Huelga}}, \bibinfo{journal}{Phys. Rev. Lett.}
  \textbf{\bibinfo{volume}{91}}, \bibinfo{pages}{067901}
  (\bibinfo{year}{2003}).

\bibitem[{\citenamefont{Feng et~al.}(2003)\citenamefont{Feng, Zhang, Li, Gong,
  and Xu}}]{feng_entanglement}
\bibinfo{author}{\bibfnamefont{X.-L.} \bibnamefont{Feng}},
  \bibinfo{author}{\bibfnamefont{Z.-M.} \bibnamefont{Zhang}},
  \bibinfo{author}{\bibfnamefont{X.-D.} \bibnamefont{Li}},
  \bibinfo{author}{\bibfnamefont{S.-Q.} \bibnamefont{Gong}}, \bibnamefont{and}
  \bibinfo{author}{\bibfnamefont{Z.-Z.} \bibnamefont{Xu}},
  \bibinfo{journal}{Phys. Rev. Lett.} \textbf{\bibinfo{volume}{90}},
  \bibinfo{pages}{217902} (\bibinfo{year}{2003}).

\bibitem[{\citenamefont{Duan and Kimble}(2003)}]{duan:_effic}
\bibinfo{author}{\bibfnamefont{L.~M.} \bibnamefont{Duan}} \bibnamefont{and}
  \bibinfo{author}{\bibfnamefont{H.~J.} \bibnamefont{Kimble}},
  \bibinfo{journal}{Phys. Rev. Lett.} \textbf{\bibinfo{volume}{90}},
  \bibinfo{pages}{253601} (\bibinfo{year}{2003}).

\bibitem[{\citenamefont{Simon and Irvine}(2003)}]{simon_entanglement}
\bibinfo{author}{\bibfnamefont{C.}~\bibnamefont{Simon}} \bibnamefont{and}
  \bibinfo{author}{\bibfnamefont{W.~T.~M.} \bibnamefont{Irvine}},
  \bibinfo{journal}{Phys. Rev. Lett.} \textbf{\bibinfo{volume}{91}},
  \bibinfo{pages}{110405} (\bibinfo{year}{2003}).

\bibitem[{\citenamefont{Clark et~al.}(2003)\citenamefont{Clark, Peng, Gu, and
  Parkins}}]{clark_entanglement}
\bibinfo{author}{\bibfnamefont{S.}~\bibnamefont{Clark}},
  \bibinfo{author}{\bibfnamefont{A.}~\bibnamefont{Peng}},
  \bibinfo{author}{\bibfnamefont{M.}~\bibnamefont{Gu}}, \bibnamefont{and}
  \bibinfo{author}{\bibfnamefont{S.}~\bibnamefont{Parkins}},
  \bibinfo{journal}{Phys. Rev. Lett.} \textbf{\bibinfo{volume}{91}},
  \bibinfo{pages}{177901} (\bibinfo{year}{2003}).

\bibitem[{\citenamefont{Zou et~al.}(2003)\citenamefont{Zou, Pahlke, and
  Mathis}}]{zou_4distant}
\bibinfo{author}{\bibfnamefont{X.~B.} \bibnamefont{Zou}},
  \bibinfo{author}{\bibfnamefont{K.}~\bibnamefont{Pahlke}}, \bibnamefont{and}
  \bibinfo{author}{\bibfnamefont{W.}~\bibnamefont{Mathis}},
  \bibinfo{journal}{Phys. Rev. A} \textbf{\bibinfo{volume}{68}},
  \bibinfo{pages}{024302} (\bibinfo{year}{2003}).

\bibitem[{\citenamefont{L\"utkenhaus et~al.}(1999)\citenamefont{L\"utkenhaus,
  Calsamiglia, and Suominen}}]{lutkenhaus}
\bibinfo{author}{\bibfnamefont{N.}~\bibnamefont{L\"utkenhaus}},
  \bibinfo{author}{\bibfnamefont{J.}~\bibnamefont{Calsamiglia}},
  \bibnamefont{and} \bibinfo{author}{\bibfnamefont{K.~A.}
  \bibnamefont{Suominen}}, \bibinfo{journal}{Phys. Rev. A}
  \textbf{\bibinfo{volume}{59}}, \bibinfo{pages}{3295} (\bibinfo{year}{1999}).

\bibitem[{\citenamefont{Vitali et~al.}(2000)\citenamefont{Vitali, Fortunato,
  and Tombesi}}]{vitali}
\bibinfo{author}{\bibfnamefont{D.}~\bibnamefont{Vitali}},
  \bibinfo{author}{\bibfnamefont{M.}~\bibnamefont{Fortunato}},
  \bibnamefont{and} \bibinfo{author}{\bibfnamefont{P.}~\bibnamefont{Tombesi}},
  \bibinfo{journal}{Phys. Rev. Lett.} \textbf{\bibinfo{volume}{85}},
  \bibinfo{pages}{445} (\bibinfo{year}{2000}).

\bibitem[{\citenamefont{Chimczak and Tana\'s}(2002)}]{chimczak02:_effect}
\bibinfo{author}{\bibfnamefont{G.}~\bibnamefont{Chimczak}} \bibnamefont{and}
  \bibinfo{author}{\bibfnamefont{R.}~\bibnamefont{Tana\'s}},
  \bibinfo{journal}{J. Opt. B} \textbf{\bibinfo{volume}{4}},
  \bibinfo{pages}{430} (\bibinfo{year}{2002}).

\bibitem[{\citenamefont{Jan\'e et~al.}(2002)\citenamefont{Jan\'e, Plenio, and
  Jonathan}}]{jane02}
\bibinfo{author}{\bibfnamefont{E.}~\bibnamefont{Jan\'e}},
  \bibinfo{author}{\bibfnamefont{M.~B.} \bibnamefont{Plenio}},
  \bibnamefont{and} \bibinfo{author}{\bibfnamefont{D.}~\bibnamefont{Jonathan}},
  \bibinfo{journal}{Phys. Rev. A} \textbf{\bibinfo{volume}{65}},
  \bibinfo{pages}{050302} (\bibinfo{year}{2002}).

\bibitem[{\citenamefont{Chimczak et~al.}()\citenamefont{Chimczak, Tana\'s, and
  Miranowicz}}]{chimczak:_entanglement_teleportation}
\bibinfo{author}{\bibfnamefont{G.}~\bibnamefont{Chimczak}},
  \bibinfo{author}{\bibfnamefont{R.}~\bibnamefont{Tana\'s}}, \bibnamefont{and}
  \bibinfo{author}{\bibfnamefont{A.}~\bibnamefont{Miranowicz}},
  \bibinfo{note}{(to be published)}.

\bibitem[{\citenamefont{Pellizzari}(1997)}]{pell}
\bibinfo{author}{\bibfnamefont{T.}~\bibnamefont{Pellizzari}},
  \bibinfo{journal}{Phys. Rev. Lett.} \textbf{\bibinfo{volume}{79}},
  \bibinfo{pages}{5242} (\bibinfo{year}{1997}).

\bibitem[{\citenamefont{Alexanian and Bose}(1995)}]{alexanian95}
\bibinfo{author}{\bibfnamefont{M.}~\bibnamefont{Alexanian}} \bibnamefont{and}
  \bibinfo{author}{\bibfnamefont{S.~K.} \bibnamefont{Bose}},
  \bibinfo{journal}{Phys. Rev. A} \textbf{\bibinfo{volume}{52}},
  \bibinfo{pages}{2218} (\bibinfo{year}{1995}).

\bibitem[{\citenamefont{Carmichael}(1999)}]{carmichaelKsiazkaMethods}
\bibinfo{author}{\bibfnamefont{H.~J.} \bibnamefont{Carmichael}},
  \emph{\bibinfo{title}{Statistical Methods in Quantum Optics 1}}
  (\bibinfo{publisher}{Springer}, \bibinfo{address}{Berlin},
  \bibinfo{year}{1999}).

\bibitem[{\citenamefont{Carmichael}(1993)}]{carmichaelksiazka}
\bibinfo{author}{\bibfnamefont{H.~J.} \bibnamefont{Carmichael}},
  \emph{\bibinfo{title}{An Open Systems Approach to Quantum Optics}}
  (\bibinfo{publisher}{Springer}, \bibinfo{address}{Berlin},
  \bibinfo{year}{1993}).

\bibitem[{\citenamefont{Plenio and Knight}(1998)}]{pleniotrajek}
\bibinfo{author}{\bibfnamefont{M.~B.} \bibnamefont{Plenio}} \bibnamefont{and}
  \bibinfo{author}{\bibfnamefont{P.~L.} \bibnamefont{Knight}},
  \bibinfo{journal}{Rev. Mod. Phys.} \textbf{\bibinfo{volume}{70}},
  \bibinfo{pages}{101} (\bibinfo{year}{1998}).

\end{thebibliography}
%\bibliographystyle{apsrev}
\end{document}